\documentclass[a4paper,11pt]{article}
\usepackage{pos}

\usepackage[textsize=small]{todonotes}

\title{Jet-like dihadron correlations in OO and Ne–Ne collisions at $\sqrt{s_{\rm NN}} = 5.36$~TeV with ALICE}

\author[a]{Neelkamal Mallick}
\author*[b]{Joonsuk Bae}

\onbehalf{on behalf of the ALICE Collaboration}

\affiliation[a]{Department of Physics, University of Jyv\"askyl\"a, P.O. Box 35, FI-40014, Jyv\"askyl\"a, Finland}
\affiliation[b]{Department of Physics, Sungkyunkwan University,\\
2066 Seobu-ro, Jangan-gu, Suwon 16419, Republic of Korea}

\emailAdd{neelkamal.mallick@cern.ch}
\emailAdd{joonsuk.bae@cern.ch}

\abstract{Heavy-ion collisions are primarily studied to investigate the properties of hot and dense primordial nuclear matter, known as the quark–gluon plasma (QGP). However, recent observations of hydrodynamic flow–like behavior in pp and p–Pb collisions at the LHC suggest the possible formation of QGP droplets even in the smallest collision systems. To bridge the multiplicity gap between small and large systems and to examine the applicability of hydrodynamic descriptions, light-ion collisions have been studied for the first time at the LHC. This contribution presents dihadron correlation measurements in OO and Ne–Ne collisions at $\sqrt{s_{\rm NN}} = 5.36$~TeV recorded with ALICE to explore possible medium effects.
The width of the near-side correlation peak has been studied as a function of charged-particle multiplicity at midrapidity, showing a multiplicity-dependent modification of the widths in various transverse momentum ($p_{\rm T}$) intervals in OO and Ne--Ne collisions relative to pp and pO collisions.
Further, the per-trigger normalized relative yields of near-side and away-side correlations, $I_{\rm AA}$, are measured in OO collisions for the first time, while in Ne–Ne collisions, the measurement is performed for the near side only. The observed suppression of the away-side associated yield in OO collisions relative to pp collisions is qualitatively consistent with jet–medium interactions, albeit with a weaker magnitude than in central Pb--Pb. However, the extent to which other effects, such as event-selection biases and cold-nuclear-matter effects, contribute to the observed suppression in the present kinematic regime remains an open question.

}

\FullConference{
13th International Conference on Hard and Electromagnetic Probes of High-Energy Nuclear Collisions \\
(Hard and Electromagnetic Probes 2026)\\
21-26 June 2026\\
Vanderbilt University, Nashville, Tennessee, USA
}

\begin{document}
\maketitle

\section{Introduction}

High-energy heavy-ion collisions at RHIC and the LHC have established the formation of a quark--gluon plasma (QGP)~\cite{Busza:2018rrf}, with anisotropic flow as a key signature of its collective expansion~\cite{Ollitrault:1992bk}. Long-range correlations, possibly of similar origin have also been observed in smaller systems, including high-multiplicity pp, pA, and light-nucleus collisions~\cite{CMS:2010ifv,ALICE:2012eyl,PHENIX:2018lia}, raising the question of the minimum system size requirement for QGP formation~\cite{Nagle:2018nvi,Brewer:2021kiv}. Jet quenching provides a complementary probe for QGP. In heavy-ion collisions, jet quenching appears as a suppression of the away-side associated yield in high-$p_{\rm T}$ triggered dihadron correlations and a redistribution of energy around the jet axis~\cite{STAR:2003pjh, ALICE:2011gpa, ALICE:2016gso, CMS:2016cvr}, whereas no such clear evidence has been established in small collision systems~\cite{ALICE:2017svf,ATLAS:2022iyq}. Light-ion collisions, such as pO, OO, and Ne--Ne, bridge the multiplicity gap between small and large systems. Recent ALICE measurements show nuclear-geometry-driven anisotropic flow in OO and Ne--Ne at $\sqrt{s_{\rm NN}}=5.36$~TeV~\cite{ALICE:2025flow}. Additionally, the suppression of high-$p_{\rm T}$ hadron yield and dijet imbalance~\cite{CMS:2025bta,ALICE:2026zck,ATLAS:2026fos} motivate more differential studies of parton energy loss in these systems.

High-$p_{\rm T}$ dihadron correlations provide a sensitive probe of jet--medium interactions. In Pb--Pb collisions, the away-side per-trigger associated yield is suppressed relative to a pp reference, pointing to medium-induced energy loss of the recoiling parton. At the same time, the near-side peak broadens with increasing event activity, consistent with a redistribution of soft fragments around the jet axis~\cite{ALICE:2016wkc,CMS:2016cvr}. The associated-yield modifications are commonly quantified through $I_{\rm AA}$,
which compares the jet-like yield in nucleus--nucleus collisions to that in minimum-bias pp collisions; values different from unity signal a modification of the fragmentation pattern relative to the vacuum. The same observable can be applied to OO and Ne--Ne collisions, which already exhibit collectivity and inclusive high-$p_{\rm T}$ hadron suppression~\cite{ALICE:2025flow,CMS:2025bta,ALICE:2026zck}, as well as a low-$p_{\rm T}$ near-side broadening similar to that seen in Pb--Pb~\cite{ALICENSPeakOONeNe}. In small systems, however, cold-nuclear-matter effects and biases related to the event-activity selection can also move $I_{\rm AA}$ away from unity~\cite{Gebhard:2024flv,Park:2025mbt}, so a careful interpretation is required. In this contribution, near- and away-side $I_{\rm AA}$ are presented for the first time in OO and Ne--Ne collisions at $\sqrt{s_{\rm NN}}=5.36$~TeV as a function of $p_{\rm T,~assoc}$, and compared with existing Pb--Pb results at $\sqrt{s_{\rm NN}}=2.76$~TeV~\cite{ALICE:2011gpa}. The results of multiplicity-dependent near-side peak broadening and model comparisons in pO, OO, and Ne--Ne collisions are already reported in Ref.~\cite{ALICENSPeakOONeNe}, hence, these are not repeated here.

\section{Experimental setup and data analysis}

This analysis uses OO and Ne--Ne data at $\sqrt{s_{\rm NN}}=5.36$~TeV from the 2025 LHC light-ion runs, together with a pp reference sample at $\sqrt{s}=5.36$~TeV, recorded with the upgraded ALICE detector in Run~3~\cite{Aamodt:2008zz,ALICE:2023udb}. Charged particles are reconstructed in the central barrel ($|\eta|<0.8$) with the second-generation Inner Tracking System (ITS2)~\cite{ALICE:2013nwm} and the GEM-based Time Projection Chamber (TPC)~\cite{ALICETPC:2020ann}, operated in a 0.5~T solenoidal field. Event activity is estimated at forward rapidity with the Fast Interaction Trigger (FIT): FT0C for OO and Ne--Ne, and FT0M (FT0A+FT0C) for pp. Minimum-bias events are selected with pileup rejection, and the primary vertex is required to satisfy $|z_{\rm vtx}|<10$~cm, following the same event-selection criteria as in the recent ALICE anisotropic-flow measurement~\cite{ALICE:2025flow}. Tracks are required to have hits in both the ITS2 and the TPC, to satisfy standard fit-quality criteria, and to pass DCA selections that suppress secondaries, in the range $1<p_{\rm T}<15$~GeV/$c$~\cite{ALICE:2025flow,ALICE-PUBLIC-2017-005}.

Dihadron correlations are measured as a function of $\Delta\eta$ and $\Delta\varphi$ for trigger--associated pairs with $p_{\rm T,~trig}>p_{\rm T,~assoc}$ at midrapidity in various multiplicity intervals. Pair-acceptance effects arising from the limited $\eta$ coverage are corrected by dividing the same-event distribution by a mixed-event background, built from five randomly chosen events in the same multiplicity class and within a common 1~cm-wide $z_{\rm vtx}$ window~\cite{Kopylov:1974th,ALICE:2016wkc}. Tracking efficiencies are obtained from PYTHIA~8 (Monash for pp~\cite{Skands:2014pea}, Angantyr for OO and Ne--Ne~\cite{Bierlich:2018xfw}) with \textsc{Geant4} detector simulation~\cite{GEANT4:2002zbu} as a function of $p_{\rm T}$, $\eta$, $z_{\rm vtx}$, and multiplicity, and are used to correct the per-trigger normalized correlation function~\cite{ALICE:2024dcr}
\begin{equation}
\label{eq:twocorr}
C(\Delta\eta,\Delta\varphi) = \frac{1}{N_{\rm trig}^{\rm corr}}\frac{{\rm d}^{2}N_{\rm pair}}{{\rm d}\Delta\eta{\rm d}\Delta\varphi}.
\end{equation}
Projecting Eq.~\eqref{eq:twocorr} onto $\Delta\varphi$ over $|\Delta\eta|<1.3$ yields $C(\Delta\varphi)$, from which the jet-like correlation is obtained as $J(\Delta\varphi)=C(\Delta\varphi)-P(\Delta\varphi)$, where $P(\Delta\varphi)=P_0\bigl(1+2\sum_{n=2,3} v_n\cos n\Delta\varphi\bigr)$. Here the ZYAM pedestal $P_0$ is the average of the four minima of $C(\Delta\varphi)$, and $v_2$ and $v_3$ are extracted with a \textit{low-multiplicity template fit} (LMTF) to the long-range projection $1.2<|\Delta\eta|<1.6$~\cite{ALICE:2025flow}. Near- and away-side yields are defined as $Y=\int J(\Delta\varphi)\,{\rm d}\Delta\varphi$ in $|\Delta\varphi|<1.3$ and $|\Delta\varphi-\pi|<1.3$, respectively. Their nuclear modification is quantified by
\begin{equation}
\label{eq:IAA}
I_{\rm AA} = \frac{Y^{\rm AA}_{\rm near/away}}{Y^{\rm pp\,({\rm MB})}_{\rm near/away}} ,
\end{equation}
where AA denotes OO or Ne--Ne, and minimum-bias pp collisions at the same nucleon--nucleon energy serve as the reference. Results are shown versus $p_{\rm T,~assoc}$ for $6<p_{\rm T,~trig}<8$ and $8~<~p_{\rm T,~trig}~<~15$~GeV/$c$, and compared with Pb--Pb $I_{\rm AA}$ at $\sqrt{s_{\rm NN}}=2.76$~TeV~\cite{ALICE:2011gpa}. Systematic uncertainties are estimated by varying the vertex and track selections, the event-activity estimator, the ZYAM pedestal, and the LMTF setup (including the long-range $\Delta\eta$ interval). Only variations exceeding $1\sigma$ by the Barlow criterion~\cite{Barlow:2002yb} are retained and combined in quadrature. 

\section{Results}

\begin{figure}[t]
    \centering
    \includegraphics[width=0.95\textwidth]{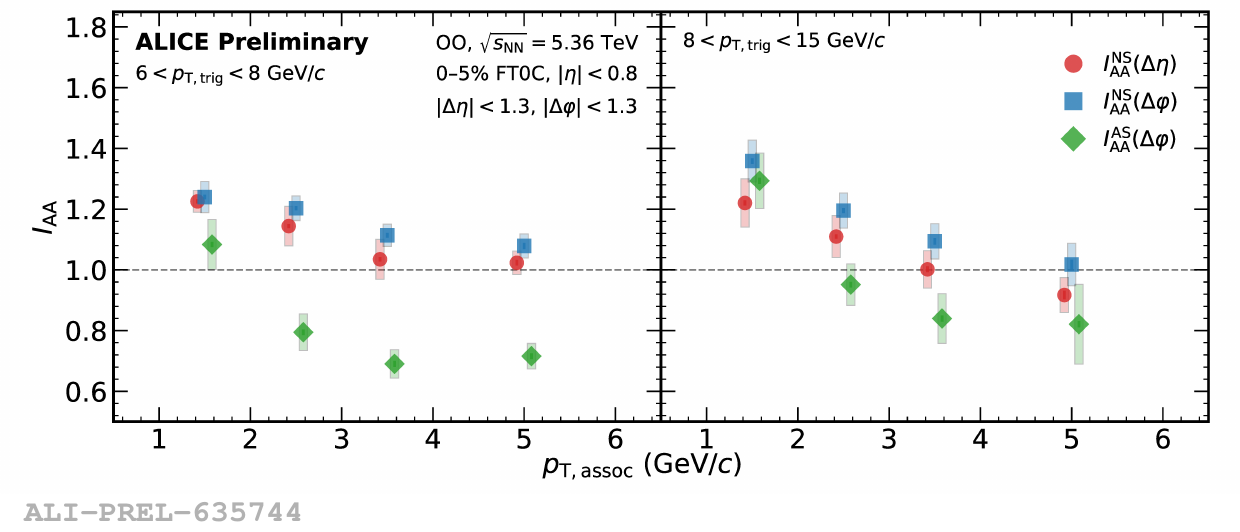}
    \caption{Near-side and away-side $I_{\rm AA}$ as a function of $p_{\rm T,~assoc}$ in 0--5\% central OO collisions at $\sqrt{s_{\rm NN}}~=~5.36$~TeV, for trigger intervals $6 < p_{\rm T,~trig} < 8$~GeV/$c$ (left) and $8 < p_{\rm T,~trig} < 15$~GeV/$c$ (right). Near-side yields are obtained from both $\Delta\eta$ and $\Delta\varphi$ projections.}
    \label{fig:iaa_overlay}
\end{figure}

\begin{figure}[t]
    \centering
    \begin{minipage}[t]{0.48\textwidth}
        \centering
        \includegraphics[width=\linewidth]{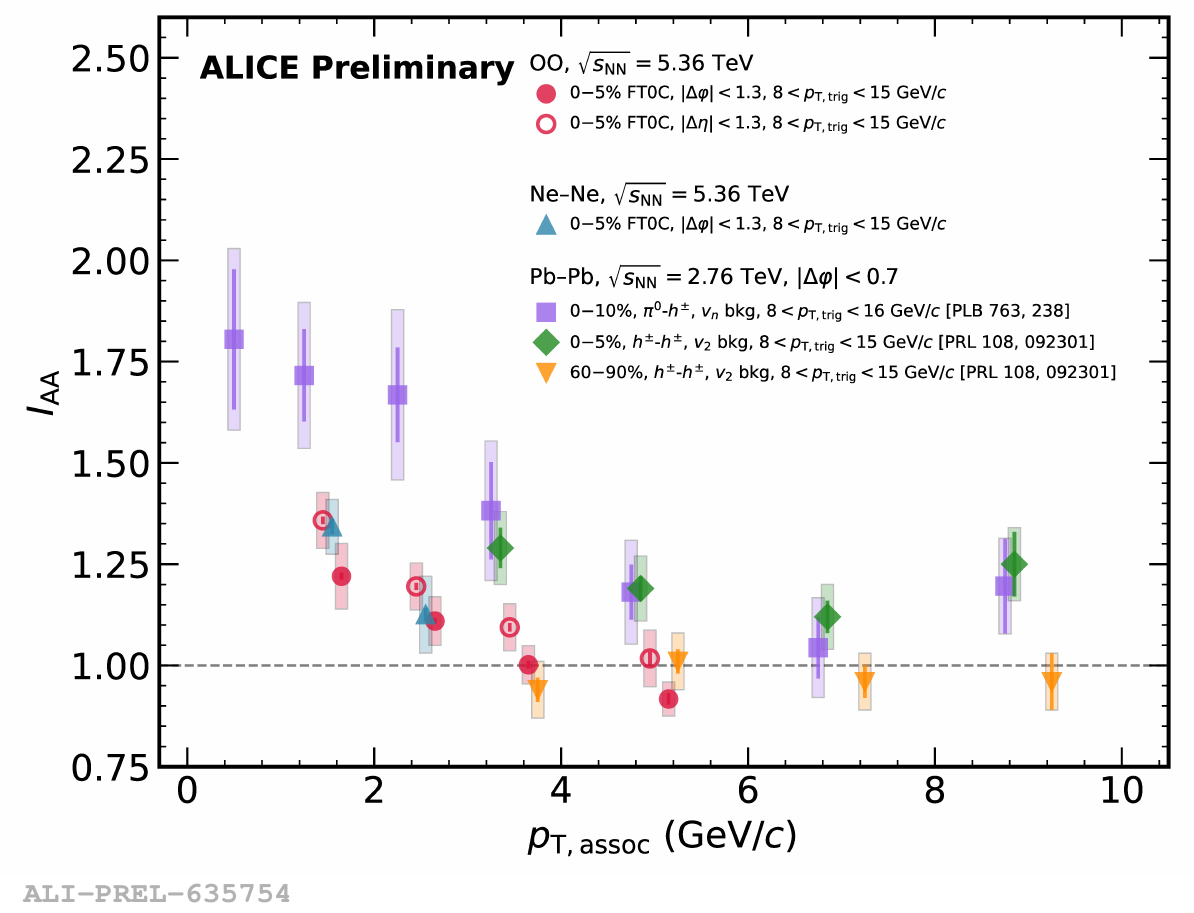}
    \end{minipage}\hfill
    \begin{minipage}[t]{0.48\textwidth}
        \centering
        \includegraphics[width=\linewidth]{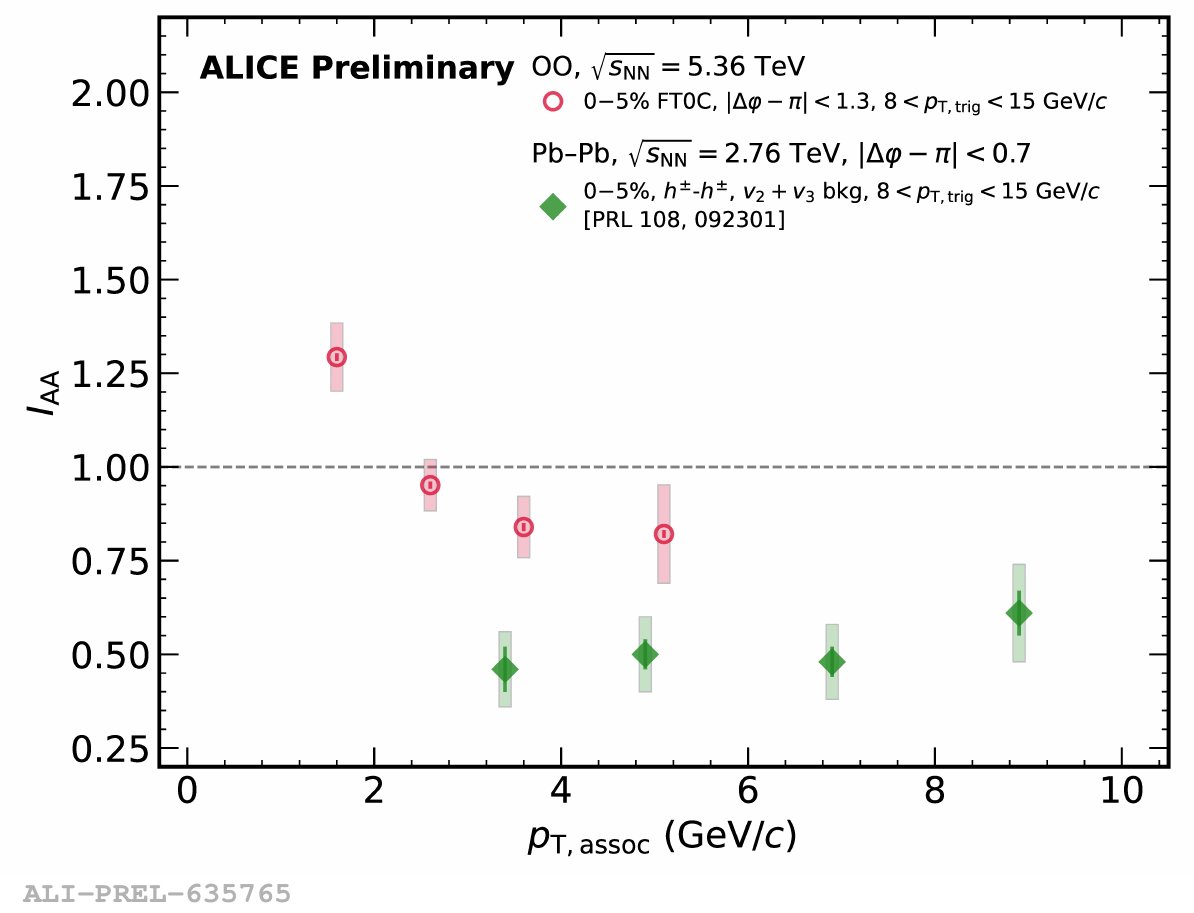}
    \end{minipage}
    \caption{Near-side (left) and away-side (right) $I_{\rm AA}$ as a function of associated-particle transverse momentum $p_{\rm T,~assoc}$ in central OO and Ne--Ne collisions at $\sqrt{s_{\rm NN}}=5.36$~TeV, compared with Pb--Pb measurements at $\sqrt{s_{\rm NN}}=2.76$~TeV. Trigger particles are selected in $8 < p_{\rm T,~trig} < 15$~GeV/$c$.}
    \label{fig:iaa_ns}
\end{figure}

Figures~\ref{fig:iaa_overlay} and~\ref{fig:iaa_ns} show near-side (NS) and away-side (AS) $I_{\rm AA}$ versus $p_{\rm T,~assoc}$ in central OO collisions at $\sqrt{s_{\rm NN}}=5.36$~TeV. At $p_{\rm T,~assoc}<2$~GeV/$c$ both sides are enhanced relative to pp; at higher $p_{\rm T,~assoc}$ the NS $I_{\rm AA}$ decreases modestly towards unity, while the AS falls below unity ($1.4\sigma$ deviation for the $8<p_{\rm T,~trig}<15\times4<p_{\rm T,~assoc}<6$~GeV/$c$ bin at $0$--$5\%$). NS yields from the $\Delta\eta$ and $\Delta\varphi$ projections agree within uncertainties, so $I_{\rm AA}^{\rm NS}$ is robust to the projection choice and not driven by a strong longitudinal--azimuthal peak asymmetry; Ne--Ne is likewise compatible with OO. Relative to Pb--Pb at $\sqrt{s_{\rm NN}}=2.76$~TeV~\cite{ALICE:2011gpa}, 60--90\% peripheral Pb--Pb is compatible with 0--5\% OO on the NS in matched $p_{\rm T}$ bins, whereas both the low-$p_{\rm T}$ NS enhancement and the high-$p_{\rm T}$ AS suppression are stronger in central Pb--Pb.

\section{Summary}

The measured $I_{\rm AA}$ in light-ion collisions shows a low-$p_{\rm T}$ enhancement and, a decreasing trend on both the near and away sides at higher $p_{\rm T}$.
The observation of high-$p_{\rm T}$ yield suppression via away-side $I_{\rm AA}<1$ in OO collisions is qualitatively consistent with jet--medium interactions, albeit with a weaker magnitude than in central Pb--Pb. Together with the similarity between central OO and peripheral Pb--Pb on the near side, these results offer a useful benchmark for models of parton energy loss and soft-fragment redistribution across system size. The potential influence of event-selection biases and cold-nuclear-matter effects, which are not addressed in this analysis, remains to be investigated. Complementary reconstructed-jet measurements will help consolidate this interpretation.

\section*{Acknowledgement}
N. M. is supported by the Academy of Finland through the Center of Excellence in Quark Matter (Grant No. 346328).
J. Bae acknowledges support from the BK21 FOUR Project of the Department of Physics, Sungkyunkwan University (No. 4120240114898), and from the Korea–CERN International Cooperation Program through the National Research Foundation of Korea (NRF), funded by the Ministry of Science and ICT (Grant No. NRF-2008-00458).


\begin{thebibliography}{99}

\bibitem{Busza:2018rrf}
W.~Busza, K.~Rajagopal and W.~van der Schee,
\emph{Heavy Ion Collisions: The Big Picture, and the Big Questions},
\href{https://doi.org/10.1146/annurev-nucl-101917-020852}
{\emph{Ann. Rev. Nucl. Part. Sci.} \textbf{68} (2018) 339}
[\href{https://arxiv.org/abs/1802.04801}{\texttt{arXiv:1802.04801 [hep-ph]}}].

\bibitem{Ollitrault:1992bk}
J.-Y.~Ollitrault,
\emph{Anisotropy as a signature of transverse collective flow},
\href{https://doi.org/10.1103/PhysRevD.46.229}
{\emph{Phys. Rev. D} \textbf{46} (1992) 229}.

\bibitem{CMS:2010ifv}
CMS Collaboration,
\emph{Observation of Long-Range Near-Side Angular Correlations in Proton-Proton Collisions at the LHC},
\href{https://doi.org/10.1007/JHEP09(2010)091}
{\emph{JHEP} \textbf{09} (2010) 091}
[\href{https://arxiv.org/abs/1009.4122}{\texttt{arXiv:1009.4122 [hep-ex]}}].

\bibitem{ALICE:2012eyl}
ALICE Collaboration,
\emph{Long-range angular correlations on the near and away side in $p$--Pb collisions at $\sqrt{s_{\rm NN}}=5.02$ TeV},
\href{https://doi.org/10.1016/j.physletb.2013.01.012}
{\emph{Phys. Lett. B} \textbf{719} (2013) 29}
[\href{https://arxiv.org/abs/1212.2001}{\texttt{arXiv:1212.2001 [nucl-ex]}}].

\bibitem{PHENIX:2018lia}
PHENIX Collaboration,
\emph{Creation of quark--gluon plasma droplets with three distinct geometries},
\href{https://doi.org/10.1038/s41567-018-0360-0}
{\emph{Nature Phys.} \textbf{15} (2019) 214}
[\href{https://arxiv.org/abs/1805.02973}{\texttt{arXiv:1805.02973 [nucl-ex]}}].

\bibitem{Nagle:2018nvi}
J.~L.~Nagle and W.~A.~Zajc,
\emph{Small System Collectivity in Relativistic Hadronic and Nuclear Collisions},
\href{https://doi.org/10.1146/annurev-nucl-101916-123209}
{\emph{Ann. Rev. Nucl. Part. Sci.} \textbf{68} (2018) 211}
[\href{https://arxiv.org/abs/1801.03477}{\texttt{arXiv:1801.03477 [nucl-ex]}}].

\bibitem{Brewer:2021kiv}
J.~Brewer, A.~Mazeliauskas and W.~van der Schee,
\emph{Opportunities of OO and $p$O collisions at the LHC},
\href{https://arxiv.org/abs/2103.01939}
{\texttt{arXiv:2103.01939 [hep-ph]}}.

\bibitem{STAR:2003pjh}
STAR Collaboration,
\emph{Evidence from $d+$Au measurements for final-state suppression of high-$p_{\rm T}$ hadrons in Au+Au collisions at RHIC},
\href{https://doi.org/10.1103/PhysRevLett.91.072304}
{\emph{Phys. Rev. Lett.} \textbf{91} (2003) 072304}
[\href{https://arxiv.org/abs/nucl-ex/0306024}{\texttt{arXiv:nucl-ex/0306024}}].

\bibitem{ALICE:2011gpa}
ALICE Collaboration,
\emph{Particle-yield modification in jet-like azimuthal di-hadron correlations in Pb--Pb collisions at $\sqrt{s_{\rm NN}}=2.76$ TeV},
\href{https://doi.org/10.1103/PhysRevLett.108.092301}
{\emph{Phys. Rev. Lett.} \textbf{108} (2012) 092301}
[\href{https://arxiv.org/abs/1110.0121}{\texttt{arXiv:1110.0121 [nucl-ex]}}].

\bibitem{ALICE:2016gso}
ALICE Collaboration,
\emph{Jet-like correlations with neutral pion triggers in pp and central Pb--Pb collisions at 2.76 TeV},
\href{https://doi.org/10.1016/j.physletb.2016.10.048}
{\emph{Phys. Lett. B} \textbf{763} (2016) 238}
[\href{https://arxiv.org/abs/1608.07201}{\texttt{arXiv:1608.07201 [nucl-ex]}}].

\bibitem{CMS:2016cvr}
CMS Collaboration,
\emph{Decomposing transverse momentum balance contributions for quenched jets in PbPb collisions at $\sqrt{s_{\rm NN}}=2.76$ TeV},
\href{https://doi.org/10.1007/JHEP11(2016)055}
{\emph{JHEP} \textbf{11} (2016) 055}
[\href{https://arxiv.org/abs/1609.02466}{\texttt{arXiv:1609.02466 [nucl-ex]}}].

\bibitem{ALICE:2017svf}
ALICE Collaboration,
\emph{Constraints on jet quenching in $p$--Pb collisions at $\sqrt{s_{\rm NN}}=5.02$ TeV measured by the event-activity dependence of semi-inclusive hadron-jet distributions},
\href{https://doi.org/10.1016/j.physletb.2018.05.059}
{\emph{Phys. Lett. B} \textbf{783} (2018) 95}
[\href{https://arxiv.org/abs/1712.05603}{\texttt{arXiv:1712.05603 [nucl-ex]}}].

\bibitem{ATLAS:2022iyq}
ATLAS Collaboration,
\emph{Strong Constraints on Jet Quenching in Centrality-Dependent $p+$Pb Collisions at 5.02 TeV from ATLAS},
\href{https://doi.org/10.1103/PhysRevLett.131.072301}
{\emph{Phys. Rev. Lett.} \textbf{131} (2023) 072301}
[\href{https://arxiv.org/abs/2206.01138}{\texttt{arXiv:2206.01138 [nucl-ex]}}].

\bibitem{ALICE:2025flow}
ALICE Collaboration,
\emph{Evidence of Nuclear Geometry-Driven Anisotropic Flow in O+O and Ne+Ne Collisions at $\sqrt{s_{\rm NN}}=5.36$ TeV},
\href{https://doi.org/10.1103/gymp-vp87}
{\emph{Phys. Rev. Lett.} \textbf{137} (2026) 082301}
[\href{https://arxiv.org/abs/2509.06428}{\texttt{arXiv:2509.06428 [nucl-ex]}}].

\bibitem{CMS:2025bta}
CMS Collaboration,
\emph{Observation of Suppressed Charged-Particle Production in Ultrarelativistic Oxygen-Oxygen Collisions},
\href{https://doi.org/10.1103/89sf-9t1x}
{\emph{Phys. Rev. Lett.} \textbf{136} (2026) 162301}
[\href{https://arxiv.org/abs/2510.09864}{\texttt{arXiv:2510.09864 [nucl-ex]}}].

\bibitem{ALICE:2026zck}
ALICE Collaboration,
\emph{Evidence for parton energy loss in oxygen--oxygen collisions at $\sqrt{s_{\rm NN}}=5.36$ TeV},
\href{https://arxiv.org/abs/2606.19967}
{\texttt{arXiv:2606.19967 [nucl-ex]}}.

\bibitem{ATLAS:2026fos}
ATLAS Collaboration,
\emph{Observation of centrality-dependent dijet transverse momentum imbalance in O+O and Ne+Ne collisions at $\sqrt{s_{\rm NN}}=5.36$ TeV with the ATLAS detector},
\href{https://arxiv.org/abs/2606.20463}
{\texttt{arXiv:2606.20463 [nucl-ex]}}.

\bibitem{ALICE:2016wkc}
ALICE Collaboration,
\emph{Anomalous evolution of the near-side jet peak shape in Pb--Pb collisions at $\sqrt{s_{\rm NN}}=2.76$ TeV},
\href{https://doi.org/10.1103/PhysRevLett.119.102301}
{\emph{Phys. Rev. Lett.} \textbf{119} (2017) 102301}
[\href{https://arxiv.org/abs/1609.06643}{\texttt{arXiv:1609.06643 [nucl-ex]}}].

\bibitem{ALICE:2018gni}
ALICE Collaboration,
\emph{Jet fragmentation transverse momentum measurements from di-hadron correlations in $\sqrt{s}=7$ TeV pp and $\sqrt{s_{\rm NN}}=5.02$ TeV $p$--Pb collisions},
\href{https://doi.org/10.1007/JHEP03(2019)169}
{\emph{JHEP} \textbf{03} (2019) 169}
[\href{https://arxiv.org/abs/1811.09742}{\texttt{arXiv:1811.09742 [nucl-ex]}}].

\bibitem{ALICE:2024dcr}
ALICE Collaboration,
\emph{Multiplicity-dependent jet modification from di-hadron correlations in pp collisions at $\sqrt{s}=13$ TeV},
\href{https://doi.org/10.1007/JHEP03(2025)194}
{\emph{JHEP} \textbf{03} (2025) 194}
[\href{https://arxiv.org/abs/2409.04501}{\texttt{arXiv:2409.04501 [nucl-ex]}}].

\bibitem{ALICENSPeakOONeNe}
S.~Prasad for the ALICE Collaboration,
\emph{Jet-like di-hadron correlations in pO, OO, and Ne--Ne collisions},
\href{https://arxiv.org/abs/2609.24237}
{\texttt{arXiv:2609.24237 [nucl-ex]}}.

\bibitem{Gebhard:2024flv}
J.~Gebhard, A.~Mazeliauskas and A.~Takacs,
\emph{No-quenching baseline for energy loss signals in oxygen-oxygen collisions},
\href{https://doi.org/10.1007/JHEP04(2025)034}
{\emph{JHEP} \textbf{04} (2025) 034}
[\href{https://arxiv.org/abs/2410.22405}{\texttt{arXiv:2410.22405 [hep-ph]}}].

\bibitem{Park:2025mbt}
J.~Park, J.~L.~Nagle, D.~V.~Perepelitsa, S.~Lim and C.~Loizides,
\emph{Selection-bias effects on high-$p_{\rm T}$ yield and correlation measurements in oxygen + oxygen collisions},
\href{https://doi.org/10.1103/yr6k-97mn}
{\emph{Phys. Rev. C} \textbf{112} (2025) 064916}
[\href{https://arxiv.org/abs/2507.03603}{\texttt{arXiv:2507.03603 [nucl-ex]}}].

\bibitem{Aamodt:2008zz}
ALICE Collaboration,
\emph{The ALICE experiment at the CERN LHC},
\href{https://doi.org/10.1088/1748-0221/3/08/S08002}
{\emph{JINST} \textbf{3} (2008) S08002}.

\bibitem{ALICE:2023udb}
ALICE Collaboration,
\emph{ALICE upgrades during the LHC Long Shutdown 2},
\href{https://doi.org/10.1088/1748-0221/19/05/P05062}
{\emph{JINST} \textbf{19} (2024) P05062}
[\href{https://arxiv.org/abs/2302.01238}{\texttt{arXiv:2302.01238 [physics.ins-det]}}].

\bibitem{ALICE:2013nwm}
ALICE Collaboration,
\emph{Technical Design Report for the Upgrade of the ALICE Inner Tracking System},
\href{https://doi.org/10.1088/0954-3899/41/8/087002}
{\emph{J. Phys. G} \textbf{41} (2014) 087002}.

\bibitem{ALICETPC:2020ann}
ALICE TPC Collaboration,
\emph{The upgrade of the ALICE TPC with GEMs and continuous readout},
\href{https://doi.org/10.1088/1748-0221/16/03/P03022}
{\emph{JINST} \textbf{16} (2021) P03022}
[\href{https://arxiv.org/abs/2012.09518}{\texttt{arXiv:2012.09518 [physics.ins-det]}}].

\bibitem{ALICE-PUBLIC-2017-005}
ALICE Collaboration,
\emph{The ALICE definition of primary particles},
\href{https://cds.cern.ch/record/2270008}
{\texttt{ALICE-PUBLIC-2017-005}} (2017).

\bibitem{Kopylov:1974th}
G.~I.~Kopylov,
\emph{Like particle correlations as a tool to study the multiple production mechanism},
\href{https://doi.org/10.1016/0370-2693(74)90263-9}
{\emph{Phys. Lett. B} \textbf{50} (1974) 472}.

\bibitem{Skands:2014pea}
P.~Skands, S.~Carrazza and J.~Rojo,
\emph{Tuning PYTHIA 8.1: the Monash 2013 Tune},
\href{https://doi.org/10.1140/epjc/s10052-014-3024-y}
{\emph{Eur. Phys. J. C} \textbf{74} (2014) 3024}
[\href{https://arxiv.org/abs/1404.5630}{\texttt{arXiv:1404.5630 [hep-ph]}}].

\bibitem{Bierlich:2018xfw}
C.~Bierlich, G.~Gustafson, L.~L{\"o}nnblad and H.~Shah,
\emph{The Angantyr model for Heavy-Ion Collisions in
\textsc{PYTHIA}8},
\href{https://doi.org/10.1007/JHEP10(2018)134}
{\emph{JHEP} \textbf{10} (2018) 134}
[\href{https://arxiv.org/abs/1806.10820}{\texttt{arXiv:1806.10820 [hep-ph]}}].

\bibitem{GEANT4:2002zbu}
S.~Agostinelli \textit{et al.},
\emph{Geant4---a simulation toolkit},
\href{https://doi.org/10.1016/S0168-9002(03)01368-8}
{\emph{Nucl. Instrum. Meth. A} \textbf{506} (2003) 250}.

\bibitem{Barlow:2002yb}
R.~Barlow,
\emph{Systematic Errors: facts and fictions},
\href{https://arxiv.org/abs/hep-ex/0207026}
{\texttt{arXiv:hep-ex/0207026}}.



\end{thebibliography}
\end{document}